\documentclass[twocolumn,showpacs,preprintnumbers,amsmath,amssymb]{revtex4}

\usepackage{graphicx}
\usepackage{dcolumn}
\usepackage{bm}
\usepackage{epsfig}

\begin{document}
\preprint{IST/CFP 6.2005-M J Pinheiro}

\title[]{The role of magnetic flux and wall current drain on anomalous diffusion}
\author{Mario J. Pinheiro}
\address{Department of Physics and Center for Plasma Physics,\&
Instituto Superior Tecnico, Av. Rovisco Pais, \& 1049-001 Lisboa,
Portugal} \email{mpinheiro@ist.utl.pt}

\thanks{We acknowledge partial financial support from Funda\c{c}\~{a}o Calouste Gulbenkian and the Rectorate of the
Technical University of Lisbon.}




\date{\today}
\begin{abstract}
A theoretical analysis of the anomalous diffusion transport
mechanism suggests a possible connection between wall current
drain and magnetic flux through the orbital trajectories of
charged particles in a plasma submitted to a strong magnetic
field. Then Bohm diffusion coefficient is straightforwardly
obtained.
\end{abstract} \maketitle

The problem of the plasma-wall interactions is of fundamental
importance in plasma physics. Studying the anomalously high
diffusion of ions across magnetic field lines in Calutron ion
sources (electromagnetic separator used by E. Lawrence for uranium
isotopes) gave the firsts indications of the onset of a new
mechanism~\cite{Bohm}. Simon advanced first with a suggestion that
the observed losses could be explained by the highly anisotropic
medium induced by the magnetic field lines, favoring electron
current to the conducting walls - the electron "short-circuit"
problem~\cite{Simon}. Experiments done by
Geissler~\cite{Geissler1} in the 1960's have shown that diffusion
in a plasma across a magnetic field was nearly classical
(standard) diffusion when insulating walls impose plasma
ambipolarity, but in the presence of conducting walls charged
particles diffuse at a much higher rate.

This problem of plasma-wall interaction becomes more complex when
a complete description is aimed of a magnetized nonisothermal
plasma transport in a conducting vessel~\cite{Beilinson}. In the
area of fusion reactors, there is strong indication that for
plasmas large but finite Bohm-like diffusion coefficient appears
above a certain range of $B$~\cite{Montgomery2}.

Progress in the understanding of the generation of confinement
states in a plasma is fundamental~\cite{Itoh} to pursue the dream
of a fusion reactor~\cite{Bickerton,Shafranov}. Anomalous
diffusion is a cornerstone in this quest, as recent research with
tokamaks suggest that the containment time $\tau \approx 10^8
R^2/2D_B$, with $R$ denoting the minor radius of a tokamak plasma
and $D_B$ is the Bohm diffusion coefficient ~\cite{Rostoker}.
Controlled nuclear fusion experiments have shown that transport of
energy and particles across magnetic field lines is anomalously
large (i.e, not predicted by classical collision theory).

The conjecture made by Bohm is that the diffusion coefficient is
$D_B=\alpha kT/eB$, where $T$ is the plasma temperature and
$\alpha$ is a numerical coefficient, empirically taken to be
$1/16$ ~\cite{Bohm}. Usually the origin of the anomalous diffusion
has been assumed to be due to the turbulence of small-scale
instabilities (see, for example,
Refs.~\cite{Montgomery2,Taylor,Montgomery1}).

In a seminal paper~\cite{Robertson} a conjecture was proposed
based on the principle of minimum entropy-production rate, stating
that a plasma will be more stable whenever the internal product of
the current density by an elementary conducting area $d
\mathcal{A}$ at every point of the boundary - excluding the
surface collecting the driving current - is null, $(\mathbf{j}
\cdot \mathcal{A})=0$.

The work reported in this Letter proposes a mechanism of wall
current drain set up together with the frozen-in effect of the
magnetic field as a possible alternative explanation of the
anomalous diffusion mechanism. From collisional low temperature
plasmas to a burning fusion plasma subject the plasma confinement
vessel to strong wall load, both in stellarator or tokamak
operating modes, this explanation could be of considerable
interest.

The general idea proposed by Robertson~\cite{Robertson} assume
that the plasma boundary is composed of small elements of area
$\mathcal{A}_i$, each one isolated from the others but each one
connected to the exterior circuit through its own resistor $R_i$
and voltage $V_i$. The entropy production rate in the external
circuits is
\begin{equation}\label{Eq1}
\frac{d S}{d t} = \sum_i \frac{1}{T} (\mathbf{j}_i \cdot
\mathcal{A}_i)^2 R_i,
\end{equation}
where $T$ is the temperature of the resistors, supposed to be in
thermal equilibrium with all the others.

In fact, a straightforward application of Eq.~\ref{Eq1} to a cold
plasma made of electrons and just one ion component gives
\begin{equation}\label{Eq2}
\frac{d S}{dt} = \frac{e^2}{T} (-n_e \mu_e \mathbf{E} + D_e \nabla
n_e + n_i \mu_i \mathbf{E} - D_i \nabla n_i)^2 \mathcal{A}^2 R.
\end{equation}
Under the usual assumptions of quasi-neutrality and
quasi-stationary plasma (see, for example, Ref. ~\cite{Roth})
\begin{equation}\label{Eq3}
\begin{array}{cc}
  \frac{n_i}{n_e}=\epsilon=const. ; & n_e \mathbf{v_e}=n_i
  \mathbf{v}_i.
  \\
\end{array}
\end{equation}
Hence, Eq.~\ref{Eq1} becomes
\begin{equation}\label{Eq4}
\frac{dS}{dt}=\frac{e^2}{T} [\mathbf{E}(\epsilon \mu_i - \mu_e)n_e
+ \nabla n_e (D_e - D_i \epsilon)]^2 \mathcal{A}^2 R.
\end{equation}
If there is no entropy production $\dot{S}=0$, and then the
ambipolar electric field is recovered~\cite{Roth}
\begin{equation}\label{amb1}
\mathbf{E}=\frac{D_e - \epsilon D_i}{\mu_e - \mu_i
\epsilon}\frac{\nabla n_e}{n_e}.
\end{equation}
This conceptual formulation provides new insight regarding
ambipolar diffusion. In a thermal equilibrium state, a plasma
confined by insulating walls will have an effective coefficient
given by the above Eq.~\ref{amb1}, a situation frequently
encountered in industrial applications. This example by itself
relates ambipolar diffusion with no entropy production in the
plasma. Allowing plasma currents to the walls, entropy production
is greatly enhanced, generating altogether instabilities and
plasma losses~\cite{Robertson}.

But confined plasmas are in a far-nonequilibrium state (with
external surroundings) and it is necessary to establish a
generalized principle that rule matter, which we develop in the
next lines.

Experiments give evidence of transport of particles and energy to
the walls~\cite{Luce}. At the end of the 1960s, experimental
results obtained in weakly ionized plasma~\cite{Geissler1} and in
a hot electron plasma~\cite{Ferrari} (this one proposing a
possible mechanism of flute instability) indicated the strong
influence conducting walls have on plasma losses across magnetic
field lines. Geissler~\cite{Geissler1} suggested that the most
probable explanation was due to the existence of diffusion-driven
current flow through the plasma to the walls. Beilinson {\it et
al.}~\cite{Beilinson} shown the possibility to control the
profiles of plasma parameters by applying potential difference to
various parts of the conducting walls. Concerning fusion
reactions, Taylor~\cite{Taylor} provided a new interpretation of
tokamak fluctuations as due to an inward particle flux resulting
from the onset of filamentary currents.

We consider a simple axisymmetric magnetic configuration with
magnetic field lines parallel to z-axis with a plasma confined
between two electrodes (see Fig.1). In general terms, a particle
motion in a plasma results in a massive flux. As long as the flux
is installed, the flux will depends naturally on a force
$\mathbf{F}$ - in this case the pressure gradient-driven process
of diffusion to the wall - responsible of the wall driven current
$\mathbf{j}$. According to the fundamental thermodynamics
relation, the plasma internal energy variation $dU$ is related to
the amount of entropy supplied or rejected and the work done by
the driven force, through the equation
\begin{equation}\label{Eq5}
\frac{dU}{dt} = (\mathbf{j} \cdot \mathcal{A})^2 R + \left(
\mathbf{F} \cdot \frac{d \mathbf{r}}{dt} \right).
\end{equation}
The last term we identify with the macroscopic diffusion velocity
$\mathbf{v}_d$ depicting the process of plasma expansion to the
wall. To simplify somehow the calculations we assume a single
plasma fluid under the action of a pressure gradient
($\mathbf{F}=\mathcal{A} L dp/dy$).

In the presence of steady and uniform magnetic field lines (this
simplifies the equations, but do not limit the applicability of
the model), the particles stream freely along them. From
magnetohydrodynamic we have a kind of generalized Ohm's law (see,
for example, Ref.~\cite{Kadomtsev1})
\begin{equation}\label{Eq6}
\nabla p = - e n \mathbf{E} - e n [\mathbf{v} \times \mathbf{B}] +
[\mathbf{j} \times \mathbf{B}] - \frac{e n \mathbf{j}}{\sigma},
\end{equation}
where $\sigma = e^2 n \tau_e/m_e$ is the electric conductivity,
with $\tau_e$ denoting the average collision time between
electrons and ions. Under the equilibrium condition
\begin{equation}\label{Eq7}
\nabla p = [\mathbf{j} \times \mathbf{B}],
\end{equation}
holds. Therefore, after inserting Eq.~\ref{Eq7} into Eq.~\ref{Eq6}
it is obtained the y component of velocity
\begin{equation}\label{Eq8}
v_y = - \frac{E_x}{B} - \frac{1}{\sigma B^2} \frac{d p}{d y}.
\end{equation}
From Eq.~\ref{Eq8} we have the classical diffusion coefficient
scaling with $1/B^2$ and thus implying a random walk of step
length $r_L$ (Larmor radius). To get the anomalous diffusion
coefficient we must consider the process of diffusion to the wall
- in the presence of an entropy source - with the combined action
of the wall current drain, as already introduced in Eq.~\ref{Eq5}.

Therefore, using the guiding center plasma model the particle
motion is made with velocity given by
\begin{equation}\label{Eq9}
\mathbf{j}=en\mathbf{v}_d=-\frac{[\nabla p \times
\mathbf{B}]}{B^2}.
\end{equation}
This equation form the base of a simplified theory of magnetic
confinement. In fact, the validity of Eq.~\ref{Eq9} is restrained
to the high magnetic field limit, when the Larmor radius is
shorter than the Debye radius.

Considering motion along only one direction perpendicular to the
wall (y-axis), it is clear that
\begin{equation}\label{Eq10}
(\mathbf{j} \cdot \mathcal{A})^2 = \frac{\mathcal{A}^2}{B^2}
\left( \frac{dp}{dy} \right)^2.
\end{equation}
If we consider a quasi-steady state plasma operation, the plasma
total energy should be sustained. Hence, $dU/dt=0$, and the power
associated with the driven pressure-gradient is just maintaining
the dissipative process of plasma losses on the wall.
Eq.~\ref{Eq5} govern the evolution of diffusion velocity. Hence,
we have
\begin{equation}\label{Eq12}
n v_d = - \frac{n R \mathcal{A}}{L} \frac{kT}{B^2} \frac{dn}{dy} =
-D_{T} \frac{dn}{dy},
\end{equation}
with $D_T$ denoting the transverse (across the magnetic field)
diffusion coefficient given by
\begin{equation}\label{Eq13}
D_T = \frac{n R \mathcal{A}}{L} \frac{kT}{B^2}.
\end{equation}
This new result coincides with the classical diffusion
coefficient~\cite{Roth} whenever $nR\mathcal{A}/L \equiv m
\nu_{ei}/e^2$, containing a dependence on collision frequency and
particle number density. Others theoretical approaches to this
problem were advanced by Bohm~\cite{Bohm}, who proposed an
empirically-driven diffusion coefficient associating plasma
oscillations as the source of the enhanced diffusion, while
Tonks~\cite{Tonks} have shown that the current density that is
present in a magnetically immobilized plasma is only generated by
the particle density gradient, not being associated with any drift
of matter. Simon electron "short-circuit"~\cite{Simon} proposes an
explanation for the different rates of diffusion electrons and
ions do experiment across the magnetic field as due to an
unbalance of currents flowing to the wall.

In the absence of collisions, the guiding centers of charged
particles behave as permanently attached to the same lines of
force. On the contrary, as a result of collisions with others
charged particles the guiding centers shift from one line of force
to another resulting in a diffusion of plasma across the field
lines. In our model, each orbit constitutes an elementary current
$I$ eventually crossing the wall.

\begin{figure}
  \includegraphics[width=3.5 in, height=4.5 in]{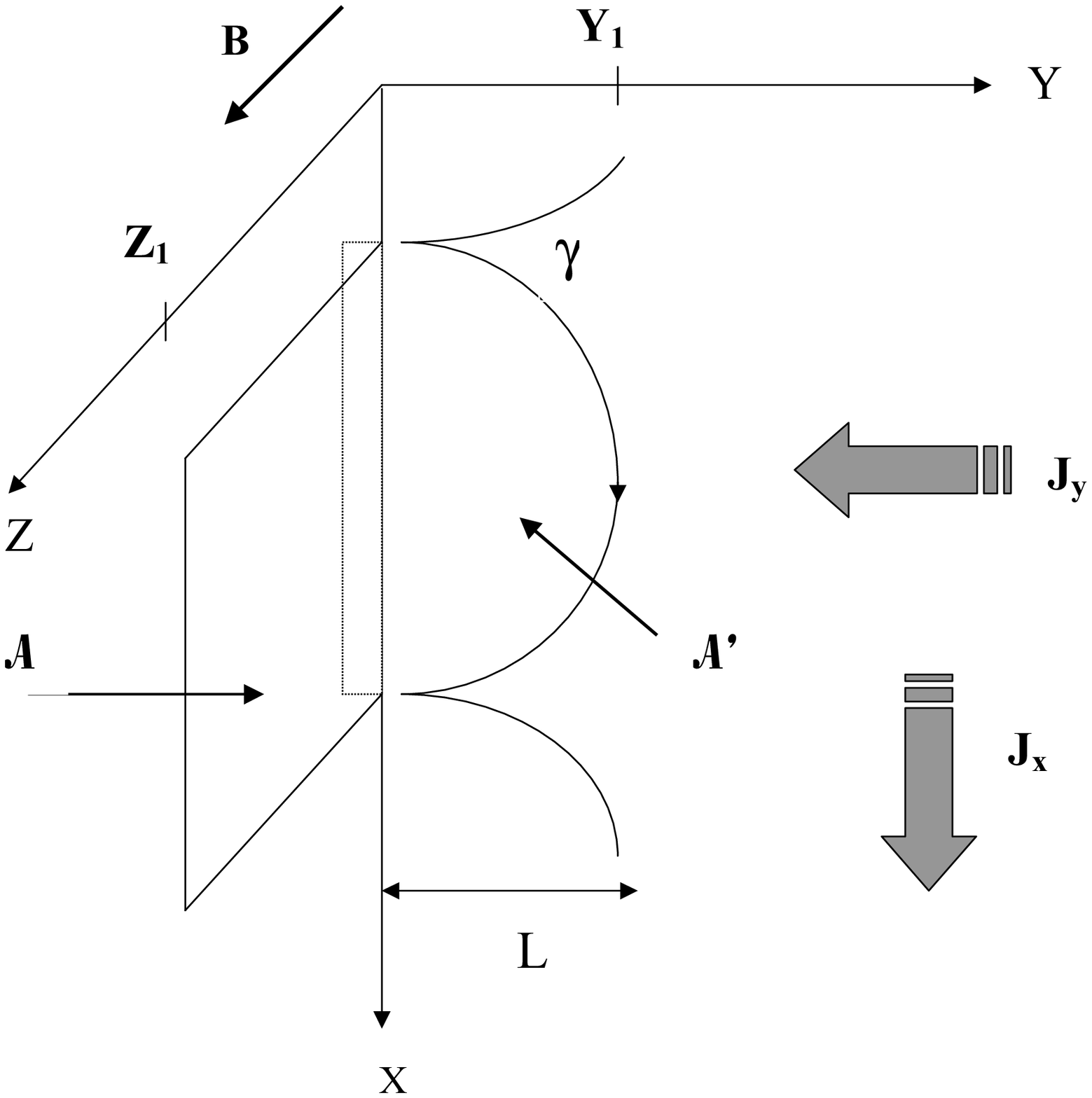}\\
  \caption{Schematic of the geometry for the plasma-wall current drain model. The uniform magnetic field
  points downward along Oz. Particles describe orbits in the plane xOy intersecting the wall
  (plan xOz). Orbits are represented by a semi-circular line for convenience.
  $L$ is the maximum distance the trajectory attains
  from the wall.}\label{fig1}
\end{figure}

However, the particle diffusion coefficient as shown in
Eq.~\ref{Eq13} gives evidence of an interplay between the
resistance the elementary circuit is submitted when in contact
with the walls in the presence of the frozen-in effect. In fact,
for sufficiently strong magnetic fields apparently a hydrodynamic
behavior of the plasma is installed ~\cite{Montgomery2,Corkum},
with the appearance of "convective cells" and the $1/B$ behavior
dominates, giving birth to the anomalous diffusion mechanism. The
onset of freezing magnetic lines is valid whenever the Lundquist
number $\mathrm{S} \gg 1$ (convection of the magnetic field
dominated medium). In this case the magnetic field lines are
frozen-in in the medium (consequence of a vortex type of character
of the magnetic field $\mathbf{B}$) and the flux of them across a
given surface is constant:
\begin{equation}\label{Eq14}
\Phi =B \mathcal{A}' = B L^2 \alpha.
\end{equation}
Remark that $\mathcal{A}'$ is now the surface delimited by the
elementary circuit $\gamma$ (see Fig.~\ref{fig1}) and $\alpha
\lesssim 1$ is just a geometrical factor (e.g. $\alpha=\pi/4$ at
the limit of a circular orbit). This situation is fundamental to
the onset of anomalous diffusion. Free electrons orbits are
helical, but as Fig.~\ref{fig1} shows, their projections at right
angles to the field are circular. Each particle orbit constitute
an elementary circuit with $B$-field cutting its surface being
associated with it an elementary flux $\Phi$. At the same time we
can envisage each orbit as constituting by itself an elementary
circuit, some of them intersecting the wall and thus the circuit
is closed inside the wall. Therefore a resistance $R$ drags the
charged flow at the conducting wall. It is therefore plausible to
associate to this elementary circuit a potential drop $V$ and the
all process being equivalent to a current $I$ flowing through the
elementary circuit.

Assuming the plasma is a typical weakly coupled, hot diffuse
plasma with a plasma parameter (number of particles in Debye
sphere) $\Lambda = n\lambda_{De}^3 \approx 1$, it is more likely
to expect nearly equal average kinetic and potential energy.
However, the typical plasma parameter encountered in glow
discharges or in nuclear fusion is $\Lambda \gg 1$. This means
that the average kinetic energy is larger than the average
potential energy. To contemplate all range of $\Lambda$ we can
relate them through the relationship
\begin{equation}\label{Eq15}
\rho V = (\mathbf{J} \cdot \mathbf{A}) \delta.
\end{equation}
Here, $\rho$ is the charge density, $\mathbf{A}$ is the vector
potential, $\mathbf{J}$ is the current density and $\delta \leq 1$
is just a parameter representing the ratio of potential to kinetic
energy. Of course, when $\Lambda \geq 1$, then $\delta \leq 1$.
This basic assumption is consistent with the hydrodynamic
approximation taken in the development of equations. The
limitations of the model are related with the unknowns $\Lambda$
and $\delta$ that can be uncovered only through a self-consistent
model of the plasma. However, our analysis of anomalous diffusion
remains general and added new insight to the phenomena.

Now suppose that the diffusion current is along y-axis
$\mathbf{J}=-J_y \mathbf{u}_y$ (see Fig.1). Consequently,
$\mathbf{A}=-A_y \mathbf{u}_y$, and then the potential drop will
depend on x-coordinate:
\begin{equation}\label{Eq16}
\rho [V(x_1) - V(x_0)] = J_y [A_y(x_1) - A_y(x_0)] \delta.
\end{equation}
Multiplying both members by the area $\mathcal{A}'=x_1 z_1$ and
length $L=y_1$, we have
\begin{equation}\label{Eq17}
Q \Delta V = I y_1 [A_y(x_1) - A_y(x_0)] \delta =I \Phi \delta.
\end{equation}
$\Phi=\oint_{\gamma} (\mathbf{A} \cdot d\mathbf{x})$ is the flux
of the magnetic field through the closed surface bounded by the
line element $d \mathbf{x}$ (elementary circuit $\gamma$, see also
Fig.\ref{fig1}). By other side, naturally, the total charge
present on the volume $\mathcal{V}=x_1 y_1 z_1$ is such as $Q=ie$,
with $i$ an integer. This integer must be related to ions charge
number. From Eq.~\ref{Eq17} we obtain
\begin{equation}\label{Eq19}
R = \frac{\Delta V}{I} = \delta \frac{\Phi }{Q} = \alpha \delta
\frac{B L^2 }{i e}.
\end{equation}
But, the particle density is given by $n=N/L\mathcal{A}$, with $N$
being now the total number of charged particles present in volume
$\mathcal{V}=\mathcal{A}L$. Since $i=N$, we retrieve finally the
so-called Bohm-diffusion coefficient
\begin{equation}\label{Eq18}
D_B = \alpha \delta \frac{kT}{eB}.
\end{equation}
However, this expression suffers of the indetermination of the
geometrical factor $\alpha$. This factor is related to the ions
charge number, it depends on the magnetic field magnitude and as
well the external operating conditions (due to increased
collisional processes, for ex.). The exact value of the product
$\alpha \delta$ can only be determined through a self-consistent
plasma model, but we should expect from the above discussion that
$\alpha \delta < 1$. Furthermore, Eq.~\ref{Eq19} can be used as a
boundary condition (simulating an electrically floating surface)
imposed when solving Poisson equation.

Also it worth to emphasize that when inserting Eq.~\ref{Eq19} into
Eq.~\ref{Eq13}, and considering the usual definition of momentum
transfer cross section, then it can be obtained a new expression
for the classical diffusion coefficient as a function of the ratio
of collisional $\nu$ and cyclotron frequency $\Omega$, although
(and in contrast with the standard expression), now also dependent
on the geometrical factor $\alpha$ and energy ratio $\delta$:
\begin{equation}\label{}
D_T = (\alpha \delta) \frac{\nu}{\Omega} \frac{kT}{m}.
\end{equation}
This explains the strong dependence of the classical diffusion
coefficient on $\nu/\Omega$ showing signs of anomalous diffusion
as discussed in Ref.~\cite{Zoran} (obtained with a time resolved
Monte Carlo simulation in an infinite gas under uniform fields)
and, in addition, the strong oscillations shown up in the
calculations of the time dependence of the transverse component of
the diffusion tensor for electrons in low-temperature rf argon
plasma. Those basic features result on one side from its
dependence on $R$, which is proportional to the flux. Therefore, a
flux variation can give an equivalent effect to the previously
proposed mechanism: whenever a decrease (or increase) in the flux
is onset through time dependence of electric and magnetic fields,
it occurs a strong increase (or decrease) of the diffusion
coefficient. By other side, when the resistance increases it
occurs a related decrease of charged particles tangential velocity
and its mean energy. So far, this model gives a new insight into
the results referred in~\cite{Zoran} and also it explains why the
same effect is not obtained from the solution of the
non-conservative Boltzmann equation as applied to an oxygen
magnetron discharges with constant electric and magnetic
fields~\cite{White}.

To summarize, we introduced in this Letter a simple mechanism
providing an interpretation of the anomalous diffusion in a
magnetized confined plasma. In fact, above a certain range of $B$
the magnetic field frozen-in effect is settled in the plasma,
implying a constant magnetic field flux through the elementary
orbits of the charge carriers. Whenever conducting walls are
bounding the plasma current drain to the walls occurs naturally
and a Bohm-like behavior of the transverse diffusion coefficient
results. The suggested mechanism could lead to a better
understanding of the mechanism of plasma-wall interaction and help
to develop a full-scale numerical modelling of present fusion
devices or collisional low-temperature plasmas.

The author would like to thank Elena Tatarova and Marques Dias for
their elucidating discussions.

\bibliographystyle{amsplain}
\bibliography{Doc2}

\end{document}